\documentclass{article}
\usepackage{
spconf,
amsmath,
graphicx,
url,
booktabs,
enumitem,
multirow,
}

\usepackage{CJKutf8}

\title{VoxHakka: A Dialectally Diverse Multi-speaker Text-to-Speech System for Taiwanese Hakka}

\name{Li-Wei Chen$^1$, Hung-Shin Lee$^1$, and Chen-Chi Chang$^2$}
\address{$^1$United Link Co., Ltd., Taiwan\\
$^2$Dept. Culture Creativity and Digital Marketing, National United University, Taiwan}

\begin{document}
\maketitle
\begin{abstract}
This paper introduces VoxHakka, a text-to-speech (TTS) system designed for Taiwanese Hakka, a critically under-resourced language spoken in Taiwan. Leveraging the YourTTS framework, VoxHakka achieves high naturalness and accuracy and low real-time factor in speech synthesis while supporting six distinct Hakka dialects. This is achieved by training the model with dialect-specific data, allowing for the generation of speaker-aware Hakka speech. To address the scarcity of publicly available Hakka speech corpora, we employed a cost-effective approach utilizing a web scraping pipeline coupled with automatic speech recognition (ASR)-based data cleaning techniques. This process ensured the acquisition of a high-quality, multi-speaker, multi-dialect dataset suitable for TTS training. Subjective listening tests conducted using comparative mean opinion scores (CMOS) demonstrate that VoxHakka significantly outperforms existing publicly available Hakka TTS systems in terms of pronunciation accuracy, tone correctness, and overall naturalness. This work represents a significant advancement in Hakka language technology and provides a valuable resource for language preservation and revitalization efforts.
\end{abstract}
\section{Introduction}
\label{sec:intro}

For low-resource languages, Text-to-Speech (TTS) systems play a crucial role in language preservation and revitalization. Speaker-aware and zero-shot TTS systems \cite{jia2018,casanova2022,wang2023,casanova2024,kim2024} contribute to preservation efforts by not only capturing the unique phonetic nuances of the language but also by enabling the reproduction of voices belonging to culturally significant individuals, even when data is scarce. Furthermore, these systems facilitate language promotion by enabling the creation of diverse audio content, including audiobooks, short films, educational materials for non-native speakers, and newscasts featuring virtual anchors. This accessibility to engaging audio content can be instrumental in raising awareness and promoting the use of low-resource languages.

\begin{CJK*}{UTF8}{bsmi}
Motivated by this need, we introduce VoxHakka, a freely available, Creative Commons (CC)-licensed, high-quality multi-speaker TTS system designed specifically for Taiwanese Hakka (臺灣客家語).

Taiwanese Hakka, a major language in Taiwan, is spoken by the Hakka community, comprising approximately 15-20\% of the total population (2-3 million people)\footnote{\url{https://en.wikipedia.org/wiki/Taiwanese_Hakka/}.}. However, despite its significant speaker base, the number of fluent speakers is dwindling, particularly among younger generations who often favor Mandarin or Taiwanese Hokkien (臺灣台語). This shift in language use has resulted in a dearth of readily accessible, open, and accurately labeled Hakka language resources.

Furthermore, Taiwanese Hakka exhibits significant dialectal variation, primarily categorized into six major dialects: Sixian (四縣), Hailu (海陸), Dapu (大埔), Raoping (饒平), Zhaoan (詔安), and Nansixian (南四縣). Sixian and Hailu are the most prevalent, accounting for 57.5\% and 35.8\% of speakers, respectively. This linguistic diversity is reflected in Hakka's rich phonology and intricate grammatical structure. The language features seven tones, with variations in number and values across dialects. Its consonant system demonstrates complexity through contrasts in voicing, aspiration, and a diverse range of vowels, including monophthongs, diphthongs, and nasalized vowels. The presence of syllable-final consonants (codas), particularly in checked-tone syllables ending in /p/, /t/, or /k/, distinguishes Hakka from other Sinitic languages. This, coupled with complex tone sandhi rules influenced by syllable combinations, contributes to the intricate prosodic patterns of the language. Importantly, each Hakka dialect exhibits unique characteristics in its tones, vowels, consonants, and tone sandhi rules, further underscoring the importance of dedicated language resources for each.

These factors—data scarcity and pronunciation complexity—pose significant challenges for developing high-quality Hakka TTS systems. Consequently, there is a dearth of robust Hakka TTS systems suitable for widespread use in academia or industry\footnote{As far as we know, only Cyberon (賽微科技) and Bronci (長問科技) published Taiwanese Hakka TTS systems, but both of the companies provide only the Sixian dialect.}. Existing systems often focus solely on the Sixian dialect and offer limited speaker diversity (typically fewer than three speakers). Furthermore, these systems lack zero-shot capabilities, which would allow users to personalize their learning experience by synthesizing speech with their own voices. Such capabilities would also be invaluable in scenarios requiring voice anonymization or de-identification.
\end{CJK*}

To address these challenges, we present a comprehensive approach to developing VoxHakka, encompassing web-based corpus collection, careful selection of a suitable TTS model architecture, and robust training procedures. VoxHakka distinguishes itself with the following key features:
\begin{enumerate}[noitemsep,leftmargin=*]
\item \textbf{Comprehensive Dialectal Coverage}: VoxHakka is capable of synthesizing speech in all six major Taiwanese Hakka dialects.
\item \textbf{Ethically Sourced and Reliable Data}: The training data is ethically sourced from publicly available resources on the internet, primarily government educational institutions and publicly funded foundations. While not available for redistribution, the data's accuracy and annotations are inherently reliable due to their origins.
\item \textbf{Zero-Shot Synthesis and Computational Efficiency}: Leveraging YourTTS technology \cite{casanova2022}, VoxHakka operates as a zero-shot model, enabling the synthesis of speech for unseen speakers and even in different languages (e.g., Mandarin). Furthermore, it offers control over speaking rate and facilitates efficient inference using only CPU resources.
\item \textbf{Open Accessibility}: VoxHakka is released under a permissive CC-BY 4.0 license\footnote{\url{https://creativecommons.org/licenses/by/4.0/}.}, granting users the freedom to utilize, modify, and share the model without restrictions.
\end{enumerate}

We make our demo page and model publicly available at \url{https://voxhakka.github.io/}.

\begin{figure}[t]
\centering
\includegraphics[width=1.0\linewidth]{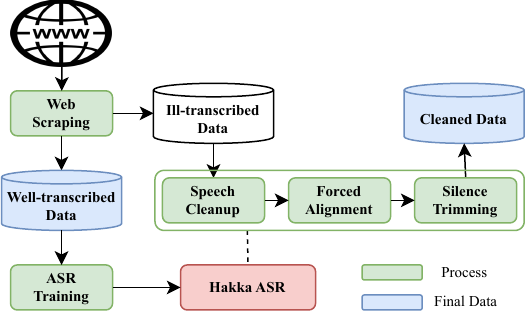}
\vspace{-15pt}
\caption{Our procedures of data acquisition and generation. The final data are used for TTS training.}
\label{fig:data}
\vspace{-10pt}
\end{figure}

\section{Acquisition of Speech for TTS}
\label{sec:format}

High-quality TTS models rely on clean, well-articulated speech data with a consistent speaking rate. We describe our methodology for acquiring and processing data from web sources to create a suitable dataset for TTS model training. Figure \ref{fig:data} illustrates this data preparation pipeline.

\begin{CJK*}{UTF8}{bsmi}
\begin{enumerate}[noitemsep,leftmargin=*]
\item \textbf{Web Scraping}: We utilize web scraping techniques to gather audio files and their corresponding transcriptions from government websites and affiliated resources, such as the Dictionary of Taiwanese Hakka\footnote{Constructed by Ministry of Education (教育部): \url{https://hakkadict.moe.edu.tw/}.}, the Bank of Hakka Examination\footnote{Constructed by Hakka Affairs Council (客委會): \url{https://elearning.hakka.gov.tw/}.}, and Hakka Radio\footnote{Funded by Hakka Public Communication Foundation (客傳會): \url{https://www.hakkaradio.org.tw/}.}. This data can be categorized into two types:
\begin{itemize}
\item Well-transcribed Data: Primarily sourced from example sentences and language learning materials, this data exhibits high transcription accuracy with strong alignment between audio and text.
\item Ill-transcribed Data: Obtained from sources like radio press releases, this data often contains discrepancies between the audio content and its accompanying text. However, the audio quality generally remains suitable for TTS training.
\end{itemize}
\item \textbf{ASR Training}: We leverage well-transcribed data, alongside a six-dialect lexicon from the Dictionary of Taiwanese Hakka, to train an Automatic Speech Recognition (ASR) system. This system's acoustic model encompasses the full range of Hakka phonemes and tones.
\item \textbf{Speech Cleanup}: Employing the trained ASR system's acoustic model (implemented using the Kaldi toolkit \cite{povey2011}), we refine the transcriptions of the ill-transcribed data. For each phrase, a biased language model is generated using its initial transcription. By adjusting discounting constants, we control the influence of this biased language model during transcription refinement. Larger constants prioritize acoustic information, yielding transcriptions closely matching the actual pronunciation.
\item \textbf{Forced Alignment}: To minimize extraneous silences and provide phoneme-level timing information, essential for certain TTS architectures, we perform forced alignment on the cleaned transcriptions using the trained ASR system. This yields precise timing information for each phoneme, including silences.
\item \textbf{Silence Trimming}: We segment phrases based on long silences (greater than 0.05 seconds), preserving up to 0.025 seconds of silence before and after each segment to ensure natural-sounding transitions.
\item \textbf{Final Data}: Finally, the well-transcribed data and the processed, cleaned data are combined to form the final dataset for training the VoxHakka TTS model.
\end{enumerate}
\end{CJK*}

\begin{figure}[t]
\centering
\includegraphics[width=1.0\linewidth]{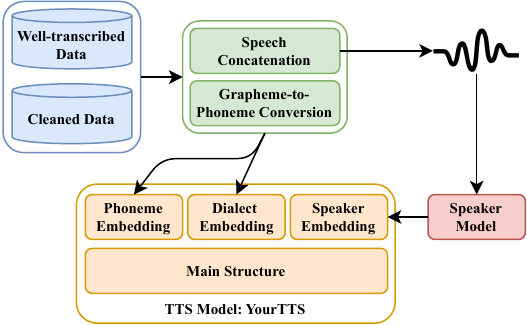}
\vspace{-15pt}
\caption{Our procedures of TTS model training. The speaker model is pre-trained.}
\label{fig:train}
\vspace{-5pt}
\end{figure}

\section{TTS Model Training}

While a well-constructed dataset is essential, further processing of text and speech data (cf. Figure \ref{fig:data}) is necessary before training the TTS model. Figure \ref{fig:train} illustrates our TTS model training pipeline.

\begin{figure}[t]
\centering
\includegraphics[width=1.0\linewidth]{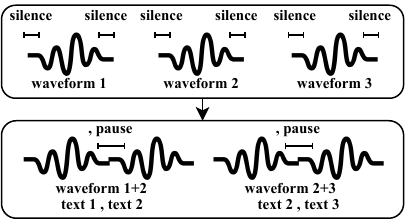}
\vspace{-15pt}
\begin{CJK*}{UTF8}{bsmi}
\caption{Diagram of speech concatenation, where waveforms 1-3 are successive segments that belong to the same utterance, and texts 1-3 are their corresponding texts. Waveform 1+2 denotes the new segment that concatenates waveform 1 and 2, while the text part becomes ``text 1,text 2'' in series. Note that the comma ``,'' represents a fixed-length pause.}
\end{CJK*}
\label{fig:concat}
\vspace{-5pt}
\end{figure}

\begin{CJK*}{UTF8}{bsmi}
\begin{enumerate}[noitemsep,leftmargin=*]
\item \textbf{Speech Concatenation}: The data preprocessing steps described previously yield short speech segments with minimal silence. To enhance the synthesis of longer, more natural-sounding sentences, and to enable the model to learn appropriate pauses between phrases (indicated by commas in text), we implement a sentence concatenation strategy. As shown in Figure \ref{fig:concat}, a long sentence is segmented into shorter waveforms (e.g., waveform 1, waveform 2, and waveform 3) with corresponding text segments (text 1, text 2, and text 3). These segments are concatenated pairwise, introducing a 0.05-second pause (0.025 seconds from each segment) at each junction. The concatenated waveforms are paired with their corresponding concatenated text, with a comma representing the pause (e.g., ``text 1, text 2'').
\item \textbf{Grapheme-to-Phoneme (G2P) Conversion}: Unlike pinyin-based writing systems, Taiwanese Hakka utilizes Chinese characters, including numerous specialized characters. Therefore, a robust G2P conversion mechanism, mapping Hakka characters to their corresponding phonetic representations, is crucial. We have developed a comprehensive G2P conversion table, leveraging the government's online Hakka dictionary, encompassing all Hakka glyphs and their phonetic transcriptions across the six major dialects. For characters with multiple pronunciations, we prioritize the most frequent variant at this stage.
\item \textbf{TTS Model Training}:  We employ YourTTS \cite{casanova2022}, a lightweight VITS-based model \cite{kim2021} incorporating language (or dialect) embedding, as our TTS system. YourTTS is well-suited for CPU deployment due to its smaller model size. Importantly, its speaker embedding mechanism eliminates the need for explicit speaker tokens during training. YourTTS accepts three kinds of inputs: phoneme embeddings derived from G2P conversion, dialect embeddings capturing pronunciation variations across dialects, and speaker embeddings generated by a separate speaker model \cite{heo2020} based on the G2P output. These speaker embeddings encode speaker-specific speech characteristics and speaking rate information.
\end{enumerate}
\end{CJK*}

\begin{CJK*}{UTF8}{bsmi}
\begin{table*}[t]
\small
\caption{Statistics of DICT, EXAM, and RADIO after web scraping.}
\vspace{5pt}
\label{tab:data_1}
\centering
\setlength{\tabcolsep}{6pt}
\begin{tabular}{lcccccccccc}
\toprule
\multirow{2}{*}{\textbf{Dialect}} & \multicolumn{3}{c}{\textbf{DICT}} & \multicolumn{3}{c}{\textbf{EXAM}} & \multicolumn{3}{c}{\textbf{RADIO}} & \multicolumn{1}{c}{\textbf{Total}}\\
\cmidrule(lr){2-4}
\cmidrule(lr){5-7}
\cmidrule(lr){8-10}
\cmidrule(lr){11-11}
& \textbf{\#Utt.} & \textbf{Hours} & \textbf{\#Char/Sec} & \textbf{\#Utt.} & \textbf{Hours} & \textbf{\#Char/Sec} & \textbf{\#Utt.} & \textbf{Hours} & \textbf{\#Char/Sec} & \textbf{Hours}\\
\toprule
Sixian (四縣) & 15,221 & 4.27 & 2.20 & 2,729 & 3.73 & 2.47 & 1,180 & 57.15 & 3.94 & 65.15\\ 
Hailu (海陸) & 15,322 & 5.17 & 1.82 & 4,986 & 6.06 & 3.10 & 915 & 43.54 & 3.90 & 54.77\\ 
Dapu (大埔) & 14,363 & 7.60 & 1.13 & 4,980 & 6.79 & 2.66 & 256 & 12.05 & 3.54 & 26.44\\ 
Raoping (饒平) & 14,258 & 3.50 & 2.54 & 3,171 & 8.94 & 1.19 & 68 & 2.75 & 3.50 & 15.19\\ 
Zhaoan (詔安) & 11,245 & 3.55 & 1.70 & 4,973 & 6.82 & 2.67 & 64 & 2.79 & 2.91& 13.16\\ 
Nansixian (南四縣) & 14,698 & 5.72 & 1.55 & - & - & - & - & - & - & 5.72\\
\midrule
\bf Total & 85,107 & 29.81 & 1.72 & 20,839 & 32.34 & 2.32 & 2,483 & 118.28 & 3.85 & 180.43\\
\bottomrule
\end{tabular}
\vspace{-5pt}
\end{table*}
\end{CJK*}

\begin{CJK*}{UTF8}{bsmi}
\begin{table*}[t]
\small
\caption{Statistics of DICT, EXAM, and RADIO after speech cleanup.}
\vspace{5pt}
\label{tab:data_2}
\centering
\setlength{\tabcolsep}{6pt}
\begin{tabular}{lcccccccccc}
\toprule
\multirow{2}{*}{\textbf{Dialect}} & \multicolumn{3}{c}{\textbf{DICT}} & \multicolumn{3}{c}{\textbf{EXAM}} & \multicolumn{3}{c}{\textbf{RADIO}} & \multicolumn{1}{c}{\textbf{Total}}\\
\cmidrule(lr){2-4}
\cmidrule(lr){5-7}
\cmidrule(lr){8-10}
\cmidrule(lr){11-11}
& \textbf{\#Utt.} & \textbf{Hours} & \textbf{\#Char/Sec} & \textbf{\#Utt.} & \textbf{Hours} & \textbf{\#Char/Sec} & \textbf{\#Utt.} & \textbf{Hours} & \textbf{\#Char/Sec} & \textbf{Hours}\\
\toprule
Sixian (四縣) & 14,331 & 3.75 & 6.89 & 5,126 & 2.52 & 6.26 & 60,639 & 44.74 & 6.63 & 51.01\\ 
Hailu (海陸) & 14,415 & 4.68 & 5.08 & 11,136 & 5.36 & 6.39 & 54,061 & 33.72 & 7.04 & 43.76\\ 
Dapu (大埔) & 13,386 & 7.00 & 2.96 & 12,386 & 5.77 & 6.00 & 14,686 & 8.89 & 7.19 & 21.66\\ 
Raoping (饒平) & 11,582 & 3.00 & 5.44 & 6,463 & 3.19 & 6.03 & 4,406 & 2.15 & 7.75 & 8.34\\ 
Zhaoan (詔安) & 9,961 & 2.80 & 6.94 & 12,913 & 5.72 & 6.31 & 4,389 & 2.06 & 6.92 & 10.58\\ 
Nansixian (南四縣) & 13,781 & 4.96 & 5.50 & - & - & - & - & - & - & 4.96\\
\midrule
\bf Total & 77,456 & 26.19 & 5.09 & 48,024 & 22.56 & 6.20 & 138,181 & 91.56 & 6.87 & 140.31\\
\bottomrule
\end{tabular}
\end{table*}
\end{CJK*}

\section{Experiments}

\subsection{Scraped and Cleaned Data}

Our data primarily originates from three publicly available resources on the Internet: the Dictionary of Taiwanese Hakka (DICT), the Bank of Hakka Examination Examples (EXAM), and Hakka Radio (RADIO). DICT and EXAM primarily consist of well-transcribed complete sentences. In contrast, RADIO's transcriptions often exhibit inconsistencies with the spoken content, sometimes including Mandarin pronunciations instead of Hakka. We address these inconsistencies using our speech cleanup techniques. Tables \ref{tab:data_1} and \ref{tab:data_2} provide detailed statistics for these datasets.

As evident from the tables, we collected a total of 180.53 hours of transcribed Hakka speech from online sources. RADIO constitutes the largest contributor, providing over half of the data for both the Sixian and Hailu dialects. While RADIO contains fewer individual utterances compared to DICT and EXAM, each utterance represents a complete news report, averaging around 3 minutes in length—significantly longer than the sentences in DICT and EXAM. After applying speech cleanup and silence trimming, 77.72\% of the scraped speech is deemed suitable for TTS training. Silence trimming also leads to a significant increase in the number of characters per second (\#Char/Sec), a metric reflecting speaking rate.

We observe a substantial disparity in the amount of speech data available for Dapu, Raoping, Zhaoan, and Nansixian dialects compared to Sixian and Hailu. While this discrepancy may partly reflect differences in speaker population size, it underscores the pressing need for corpus preservation and revitalization efforts for low-resource languages and dialects. However, the continuous growth of RADIO's news archive provides an encouraging source of high-quality speech data. Leveraging these expanding resources will be crucial for future improvements and updates to the VoxHakka TTS model.

\subsection{TTS Model Structure of VoxHakka}

Following the YourTTS architecture, we incorporate 4-dimensional trainable language embeddings into the input character embeddings and sum the external speaker embeddings with the text encoder outputs and decoder outputs, which are then sent to the duration predictor and vocoder. 
Similar to the VITS model, our decoder comprises four stacked affine coupling layers, each consisting of four WaveNet residual blocks \cite{oord2016}. For vocoding, we employ HiFi-GAN version 1 \cite{kong2020} with the discriminator modifications proposed by Kim \textit{et al.} \cite{kim2021}.

To enable efficient end-to-end training, we connect the TTS model and vocoder using a variational autoencoder (VAE) \cite{kingma2014} with a posterior encoder composed of 16 non-causal WaveNet residual blocks.

For model training and HiFi-GAN discriminator optimization, we use the AdamW optimizer \cite{loshchilov2019} with betas of 0.8 and 0.99, weight decay of 0.01, and an initial learning rate of 0.0002, decaying exponentially by a gamma of 0.999875.

Our model hyperparameters largely align with those in the original YourTTS paper. However, we utilize eight transformer encoder layers and omit the Speaker Consistency Loss (SCL). Model training is conducted using four NVIDIA L40S GPUs, with a batch size of 54 per GPU.

\subsection{Evaluation}

To ensure a fair and comprehensive evaluation of pronunciation accuracy and intonation, we collaborated with a Hakka linguistics expert to design 15 test sentences in Hakka characters (Table \ref{tab:text}). These sentences were specifically crafted to highlight potential strengths and weaknesses in TTS systems. As a benchmark for natural pronunciation, we recorded a female professional Hakka broadcaster reading these sentences.

Despite VoxHakka's ability to synthesise the speech of six dialects of Hakka, there are no publicly available TTS systems other than the Sixian dialect for comparative evaluation. Therefore, as far as subjective listening is concerned, we only consider the Sixian dialect as a dialectal target for the evaluation. However, in our Demo Page, we still show the synthesis results of five dialects other than Sixian for the reader's reference or future comparison.

\subsubsection{Systems to be Compared}

We compared our VoxHakka system with two commercially available Hakka TTS systems: Cyberon and Bronci. Both systems focus exclusively on the Sixian dialect and provide limited information regarding their model architectures and training corpora.

Cyberon's system requires input in Hakka Pinyin (Taiwanese Hakka Romanization System) and offers adjustable speaking rate. However, it offers only a single female speaker. We utilized our G2P system to convert the 15 test sentences into Hakka Pinyin for input to Cyberon's system.

Bronci's system accepts Hakka characters directly. Based on the provided female speaker's voice characteristics, we hypothesize that this system might have utilized the Hakka Across Taiwan (HAT) corpus \cite{liao2023} for training, as the speaker's voice closely resembles that of a female speaker in the HAT corpus. However, it is essential to note that the HAT corpus is not freely available\footnote{\url{https://www.aclclp.org.tw/use_mat_c.php#hat}}.

\begin{CJK*}{UTF8}{bsmi}
\begin{table}[t]
\small
\caption{The 15 sentences in Hakka characters for CMOS.}
\vspace{5pt}
\label{tab:text}
\centering
\setlength{\tabcolsep}{2pt}
\begin{tabular}{rp{8cm}}
\toprule
1. & 吾先生著新衫，去臺北市个國家音樂廳，聽音樂會。\\
2. & 客家族群个六堆運動會會緊辦等下去，為臺灣个體育史寫下特別个一頁。\\
3. & 新竹在舊年有二十三日超過三十六度，毋知暖化个普遍同麼个有關係？\\
4. & 大家恬聲等先生解釋，收成毋好無飯好食个時節愛仰結煞？\\
5. & 歸條路吊等長長个花燈，祈求風調雨順，歸屋下人个心願，親像花燈下燒暖个光華。\\
6. & 隔壁阿伯食酒醉，騎車仔無注意椊下去圳溝肚，聽人講當嚴重。\\
7. & 叔姆屋下窗門項个塵灰當賁，愛撥畀淨來啊。\\
8. & 雨落忒多做大水，無半息雨做天旱，實在已難耐。\\
9. & 為著愛達到目標，佢輒輒噩夜，愛想辦法提高朳仔个產量。\\
10. & 牛眼拗毋著，同樹椏仔墜下來較好拗。\\
11. & 天時毋好溼氣重，樹仔當遽歿忒。\\
12. & 電鑊仔个電線係燒忒，該斯危險哩。\\
13. & 佢生來矮頓矮頓，又專門講該阿里不答个話，敢聽得？\\
14. & 食到肚扐扐仔个阿叔在藤椅項坐等啄目睡。\\
15. & 先去坐公車，下車過行斑馬線，就到了。\\
\bottomrule
\end{tabular}
\vspace{-5pt}
\end{table}
\end{CJK*}

\subsubsection{Our Synthesized Utterances}

To emphasize VoxHakka's capabilities in zero-shot settings, we opted not to synthesize speech using speaker characteristics present in our training corpus. Instead, we adopted two approaches:
\begin{enumerate}[noitemsep,leftmargin=*]
\item We synthesized a female voice in Mandarin (not Hakka) using Google's Text-to-Speech AI.
\item We recorded a female speaker uttering the test sentences in the Sixian Hakka dialect.
\end{enumerate}

Crucially, both these speaker profiles are distinct from any speakers present in our training data.

\subsubsection{Comparative Mean Opinion Scores}

We conducted a listening test employing Comparative Mean Opinion Scores (CMOS) to evaluate listener preferences across three dimensions: naturalness, pronunciation accuracy, and tone correctness. CMOS values range from -2 (system significantly worse than human speech) to +2 (system significantly better than human speech) on a continuous scale. Sixteen listeners participated in the evaluation, each answering 60 ``Which is better?'' questions (15 test sentences × 4 system comparisons = 60 questions) per evaluation aspect, totaling 180 questions per listener.

\begin{table}[t]
\small
\caption{The 15 sentences in Hakka characters for CMOS. Note that the test utterances of VoxHakka are zero-shot synthesized, and their corresponding target speakers did not appear in the training data.}
\vspace{5pt}
\label{tab:results}
\centering
\setlength{\tabcolsep}{3pt}
\begin{tabular}{cccc}
\toprule
System & Naturalness & Pronunciation Acc. & Tone Corr.\\
\toprule
Human & 0 & 0 & 0\\
Cyberon & -1.54 & -1.7 & -1.6\\
Bronci & -1.60 & \bf -1.37 & -1.62\\
\midrule
\bf VoxHakka & & & \\
Mandarin Spk Emb. & -1.05 & -1.43 & \bf -1.2\\
Hakka Spk Emb. & \bf -0.91 & -1.5 & \bf -1.2\\
\bottomrule
\end{tabular}
\vspace{-5pt}
\end{table}

\subsubsection{Results}

Table \ref{tab:results} presents a comparative evaluation of the different TTS systems for the Sixian dialect of Taiwanese Hakka, focusing on Naturalness, Pronunciation Accuracy, and tone correctness. Human speech serves as the baseline, assigned a score of 0 for all metrics. The two commercial systems, Cyberon and Bronci, demonstrate relatively lower scores, ranging from -1.37 to -1.7, indicating a perceptual gap compared to human speech across all three dimensions.

Our proposed VoxHakka system consistently outperforms both commercial systems. Employing speaker embeddings derived from both Mandarin and Hakka speech, VoxHakka achieves competitive scores. Notably, the VoxHakka (Hakka Spk Emb.) configuration, utilizing Hakka-specific speaker embeddings, attains the highest naturalness score (-0.91) among all systems, surpassing even the Mandarin-based VoxHakka variant. This finding highlights the effectiveness of language-specific speaker embeddings in capturing the unique nuances of Hakka speech.

Despite its overall strong performance, both VoxHakka versions exhibit slightly lower pronunciation accuracy scores (-1.43 and -1.5) compared to Cyberon. This suggests a potential avenue for future improvement, specifically targeting the system's acoustic modeling capabilities.

In conclusion, VoxHakka, particularly when utilizing Hakka speaker embeddings, demonstrates promising results, showcasing superior naturalness and comparable tone correctness to human speech. Future research directions could prioritize enhancing pronunciation accuracy to further minimize the gap between VoxHakka and human-quality Hakka speech synthesis.

\section{Conclusion}

We have presented VoxHakka, a freely available, CC-BY 4.0 licensed multi-speaker TTS system for all six major dialects of Taiwanese Hakka. Addressing the challenges of data scarcity and pronunciation complexity, we developed a comprehensive system encompassing web-based data collection and processing, a robust G2P conversion mechanism, and a carefully tuned TTS model based on YourTTS. Evaluations using CMOS testing demonstrate VoxHakka's superior performance compared to commercially available Hakka TTS systems, particularly in terms of naturalness. Importantly, VoxHakka's zero-shot capabilities, enabled by language-specific speaker embeddings, open new possibilities for personalized and adaptable Hakka speech synthesis.

Future research will focus on refining VoxHakka's pronunciation accuracy by exploring improvements in acoustic modeling. Additionally, we aim to expand the system's capabilities by incorporating emotional expressiveness and prosodic control. By making VoxHakka openly accessible, we hope to facilitate research, educational initiatives, and creative applications that contribute to the preservation and revitalization of Taiwanese Hakka.

\section{Acknowledgments}
\begin{CJK*}{UTF8}{bsmi}
This study was supported by Hakka language teachers Li-Fen Huang (黃麗芬) and Fon-Siin Qi (徐煥昇), whose guidance was crucial. Thanks also to Hakka Radio (講客廣播電臺) and Hakka Public Communication Foundation (客傳會) for providing valuable resources that helped complete this work.
\end{CJK*}

\bibliographystyle{IEEEbib}
\bibliography{references}

\end{document}